# Antiferromagnetic Order in the Rare Earth Halide Perovskites CsEuBr$_3$ and CsEuCl$_3$

Daniel B. Straus,[1]* Tomasz Klimczuk[2,3], Xianghan Xu,[1] and Robert J. Cava[1]*

[1]Department of Chemistry, Princeton University, Princeton, NJ 08544 USA
[2]Faculty of Applied Physics and Mathematics, Gdansk University of Technology, Narutowicza 11/12, 80-952, Gdansk, Poland
[3]Advanced Materials Centre, Gdansk University of Technology, Narutowicza 11/12, 80-952, Gdansk, Poland

*Authors to whom correspondence should be addressed. Email: dstraus@princeton.edu, rcava@princeton.edu

**Abstract**

Bulk CsEuBr$_3$ and CsEuCl$_3$ are experimentally shown to be magnetic semiconductors that order antiferromagnetically at Neél temperatures of 2.0 K and 1.0 K respectively. Given that nanoparticles and thin films of CsEuCl$_3$ have been reported to order ferromagnetically at a similar temperature, our observation of antiferromagnetic ordering in CsEuBr$_3$ and CsEuCl$_3$ expands the possible applications of halide perovskites to now include spintronic devices where both ferromagnetic and antiferromagnetic devices can be fabricated from a single material. The conclusion that CsEuCl$_3$ can be used as a switchable magnetic material is also supported by our density-functional theory calculations.

**TOC Graphic**

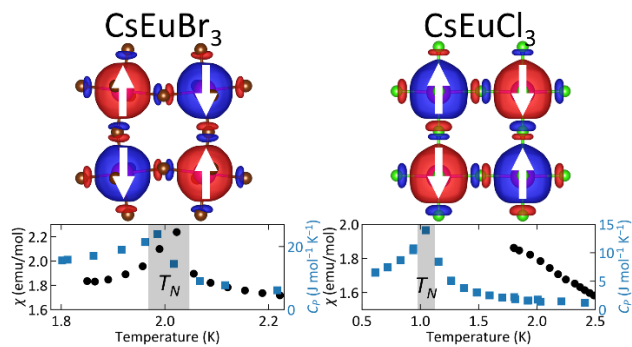

**Introduction**

Halide perovskites with the classical perovskite crystal structure are composed of a network of corner-sharing metal-halide octahedra, with the interstitial space occupied by a cation.[1,2] The choice of metal and cation is limited by size—if the cations are too large or too small, a halide perovskite cannot form.[3,4] Charge neutrality must also be obeyed—the metal is almost always in the 2+ oxidation state and the cation singly charged. Most of the metals that can form a halide perovskite are closed-shell, and magnetic ordering cannot occur because there are no unpaired spins.[5] The most studied halide perovskites contain the closed-shell metals $Pb^{2+}$ or $Sn^{2+}$ because these materials have very long carrier diffusion lengths and high absorption cross-sections,[6] allowing them to be used in high-performance solar cells with efficiencies rivaling commercial silicon-based solar cells.[7,8]

The spotlight on quantum materials has brought a renewed focus on developing new magnetic compounds that can be harnessed for spintronic applications, such as classical computers that have storage and logic in a single computational unit as well as quantum computers.[9–15] The magnetic $KMF_3$ (M = $Mn^{2+}$, $Fe^{2+}$, $Co^{2+}$, $Ni^{2+}$, or $Cu^{2+}$) perovskites were synthesized in the 1960s and all order antiferromagnetically.[16,17] $KMnCl_3$ is another antiferromagnetic halide perovskite,[18] but it is not stable and spontaneously converts into a non-perovskite phase like the halide perovskite $CsPbI_3$.[19,20] There are very few magnetic chloride, bromide, and iodide halide perovskites because open-shell 2+ transition metals are too small to form a stable perovskite, though some can be incorporated into magnetic layered two-dimensional organic-inorganic hybrid halide perovskites because the reduced dimensionality relaxes some structural constraints.[5,21]

Rare earth ions provide another way to form three-dimensional magnetic halide perovskites. Specifically, $Eu^{2+}$ has an unpaired electron in each of its seven $4f$ orbitals, allowing for magnetic interactions between $Eu^{2+}$ ions. The perovskites $CsEuX_3$ (X= Cl, Br, I) have been synthesized,[22–24] but reports of their properties are sparse. $CsEuCl_3$ and $CsEuBr_3$ have published crystal structures, but only the lattice constants of $CsEuI_3$ have been reported. The temperature-dependent magnetic susceptibility of $CsEuBr_3$ was measured above 5 K with no observed magnetic transitions, though its Weiss temperature of -2.4 K suggests that antiferromagnetic interactions are present.[23] The magnetism of bulk $CsEuCl_3$ and $CsEuI_3$ have not been reported. Thin films and nanocrystals of $CsEuCl_3$ were recently reported to be ferromagnetic with Curie temperatures of 2.5-3.0 K,[25] but the magnetism of nanocrystalline and thin film samples can differ from bulk samples due to substrate-induced strain or surface reconstruction and large surface-to-volume ratios resulting in multiple metal oxidation states.[26,27]

We demonstrate that $CsEuBr_3$ and $CsEuCl_3$ are magnetic semiconductors that order antiferromagnetically at Néel temperatures $T_N$ of 2.0 K and 1.0 K respectively. Heat capacity measurements confirm the presence of a bulk phase transition at these temperatures.[28] Surprisingly, the magnitude of the heat capacity anomalies indicate that while $CsEuBr_3$ exhibits conventional equal-moment antiferromagnetism, $CsEuCl_3$ instead exhibits amplitude-modulated antiferromagnetism where the magnetic moment periodically and incommensurately varies within the structure.[28] Antiferromagnetism is in these materials is consistent with the Goodenough-Kanamori rule for ~180° magnetic superexchange interactions through an anion with filled valence orbitals,[29] and density-functional theory calculations confirm an antiferromagnetic configuration is slightly more stable than a ferromagnetic configuration. This is the first report of antiferromagnetic ordering in a pure-phase halide perovskite with a rare earth element on the B-site, demonstrating that rare earth halide perovskites can be considered for use in spintronic devices. Furthermore, our finding of antiferromagnetism in bulk $CsEuCl_3$ in contrast to the previous report of ferromagnetism in nanocrystals and thin films[25] shows that confinement and/or strain can be used to tune magnetic interactions in halide perovskites from antiferromagnetic to ferromagnetic. This conclusion is also supported by our calculations because both the antiferromagnetic and ferromagnetic configurations of CsEuCl$_3$ are more stable than a hypothetical nonmagnetic configuration.[26,27]

**Results and Discussion**

Figure 1A shows powder X-ray diffraction patterns of CsEuBr$_3$ (black) and CsEuCl$_3$ (red), with Rietveld refinements in grey. CsEuBr$_3$ was previously reported to be an orthorhombic (Pnma, #62) at room temperature with 155.97(7)° and 162.09(10)° interoctahedral Eu-Br-Eu bond angles (Figure 1B),[23] and a Rietveld refinement of our pattern (*wR*=7.0%, GoF=1.61) to this model supports this assignment. There is a small amount (0.6% w/w) of CsBr in our material.

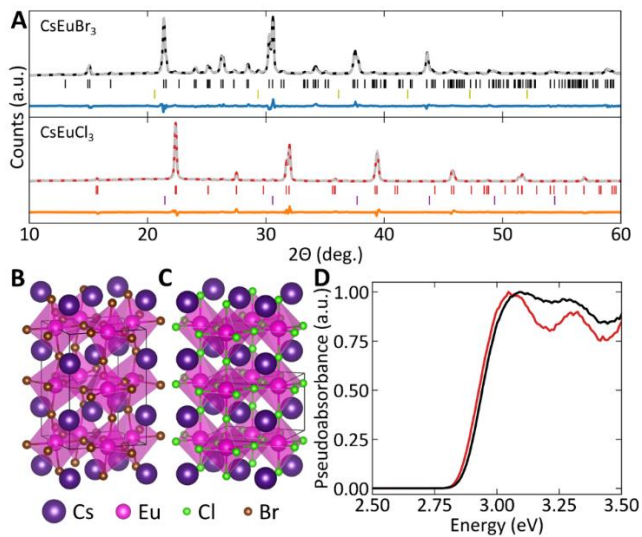

**Figure 1: Structural and optical characterization.** A) Powder diffraction patterns of CsEuBr$_3$ (black) and CsEuCl$_3$ (red) with Rietveld refinements (grey) and difference (blue, orange). Ticks indicate location of reflections of CsEuBr$_3$ (black), CsEuCl$_3$ (red), CsBr (yellow), and CsCl (purple). Depiction of structures of B) CsEuBr$_3$ and C) CsEuCl$_3$. The structure of CsEuBr$_3$ is from ref 23. D) Pseudoabsorption spectra of CsEuBr$_3$ (black) and CsEuCl$_3$ (red).

CsEuCl$_3$ is a much less distorted tetragonal perovskite (Figures 1C and S1) than CsEuBr$_3$ because of the smaller size of Cl. CsEuCl$_3$ was previously reported to be tetragonal in space group P4mm (#99) at room temperature,[22] but our powder pattern does not support this assignment because we observe additional reflections that cannot be indexed using that model. We redetermine the structure of CsEuCl$_3$ using single crystal X-ray diffraction measurements and find that CsEuCl$_3$ is best modeled in the tetragonal space group P4/mbm (#127) (Table 1). We find that CsEuCl$_3$ has 167.0(2)° and 180° interoctahedral Eu-Cl-Eu bond angles. This structural model is consistent with the reflections we observe in our powder pattern.[2] A depiction of its structure is shown in Figure 1C. A Rietveld refinement of our powder pattern (Figure 1A; wR = 8.4%, GoF=1.57) also supports this conclusion. There is a small amount (0.7% w/w) of CsCl present in our sample. There are a few very weak reflections in our powder pattern that are not indexed using the P4/mbm model but can be indexed to an orthorhombic Pnma (#62) model. However, there are many systematic absence violations if we attempt to fit our single crystal diffraction data to the orthorhombic Pnma model.

**Table 1:** Crystal data and structure refinement for CsEuCl$_3$.

| Empirical formula | CsEuCl$_3$ |
|---|---|
| Formula weight | 391.22 |
| Temperature/K | 300 |
| Crystal system | tetragonal |
| Space group | P4/mbm (#127) |
| a/Å | 7.9081(3) |
| c/Å | 5.5969(4) |
| Volume/Å$^3$ | 350.02(4) |
| Z | 2 |
| $\rho_{calc}$g/cm$^3$ | 3.712 |
| μ/mm$^{-1}$ | 15.093 |
| F(000) | 338 |
| Crystal size/mm$^3$ | 0.103 × 0.1 × 0.091 |
| Radiation | MoKα (λ = 0.71073) |
| 2Θ range for data collection/° | 7.28 to 60.85 |

| Index ranges | $-11 \leq h \leq 11$, $-11 \leq k \leq 10$, $-7 \leq l \leq 7$ |
| --- | --- |
| Reflections collected | 6143 |
| Independent reflections | 319 [$R_{int} = 0.0427$, $R_{sigma} = 0.0166$] |
| Data/restraints/parameters | 319/0/13 |
| Goodness-of-fit on $F^2$ | 1.139 |
| Final R indexes [$I>=2\sigma(I)$] | $R_1 = 0.0221$, $wR_2 = 0.0496$ |
| Final R indexes [all data] | $R_1 = 0.0332$, $wR_2 = 0.0545$ |
| Largest diff. peak/hole / e Å$^{-3}$ | 1.36/-1.74 |

CsEuBr$_3$ and CsEuCl$_3$ have optical bandgaps of 2.90(4) and 2.91(5) eV (Figures 1D and S2), which, in a tight-binding model, originate from the Eu$^{2+}$ *f-d* electronic transition,[30] explaining the nearly identical band gaps of the two compounds. Despite the atomistic origin of these states, the absorption spectra indicate that this is not a discrete, localized transition.

Both CsEuCl$_3$ and CsEuBr$_3$ are hygroscopic and degrade in the presence of moisture; this was previously reported for CsEuBr$_3$.[23] CsEuBr$_3$ also degrades when exposed to room lights, even when stored under nitrogen. It must be stored in an opaque container. CsEuCl$_3$ does not degrade under illumination.

Figure 2A shows the zero-field-cooled magnetic susceptibility ($\chi$) of CsEuBr$_3$ using a 1000 Oe field (black) with a fit (white) to the Curie-Weiss law

$$\chi = \frac{C}{T - \Theta} + \chi_0 \qquad 1$$

as well as the inverse susceptibility $(\chi - \chi_0)^{-1}$ (blue), which is linear. The Curie constant $C = 8.534(6)$ emu·K/mol, resulting in a computed magnetic moment per Eu$^{2+}$ of 8.26 $\mu_B$, which is only 4% larger than the spin-only moment for Eu$^{2+}$ ($J = 7/2$) of 7.94 $\mu_B$. The Weiss temperature $\Theta = -4.57(2)$ K, and a negative Weiss temperature indicates the dominance of antiferromagnetic coupling in the material. It is similar to a previously reported value of -2.4 K.[23] $\chi_0 = -2.50(2) \times 10^{-3}$ emu/mol.

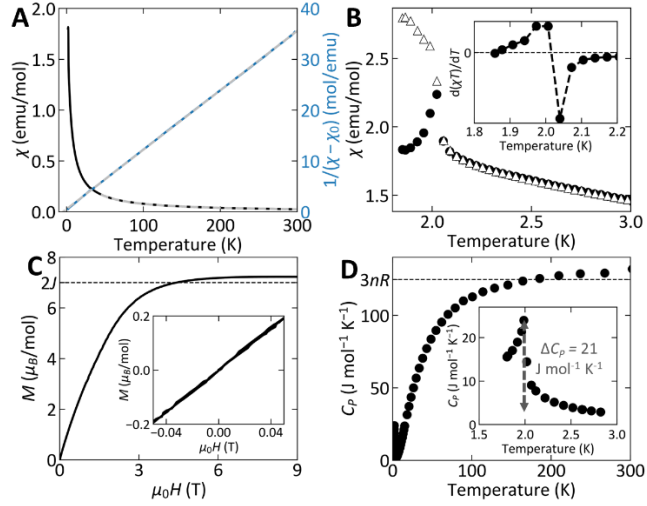

**Figure 2: CsEuBr₃ magnetism.** A) Temperature-dependent susceptibility at 1000 Oe (black) with Curie-Weiss fit (white) and inverse susceptibility (blue) with linear fit (grey). B) Zero-field-(circles) and field-cooled (triangles) susceptibility at 10 Oe. Inset: first derivative of zero-field-cooled $\chi T$. C) Field-dependent magnetization at 1.9 K. D) Constant-pressure heat capacity.

The zero-field-cooled (circles) and field-cooled (triangles) magnetic susceptibilities at a field of 10 Oe are plotted in Figure 2B. There is a discontinuity in the zero field-cooled susceptibility (inset, Figure 2B) at 2.0 K, and the downturn in $\chi$ combined with its low value suggests the onset of antiferromagnetic ordering at a Neél temperature $T_N = 2.0$ K. The onset of antiferromagnetic ordering was not observed previously because the susceptibility was only measured above 5 K.[23] There is no hysteresis in the field-dependent magnetization collected at 1.9 K (Figure 2C), further suggesting the presence of antiferromagnetic ordering. The field-dependent magnetization at 1.9 K saturates at $\mu_0 H = 6$ T because the weak antiferromagnetic interactions between spins are overcome by the applied field.

The second-order anomaly at 2.0 K in the heat capacity of CsEuBr₃ (Figure 2D) confirms the presence of a bulk phase transition, and the 21 J/K·mol magnitude is in line with the theoretical value of 20.14 J/K·mol for the heat capacity anomaly upon the transition to equal-moment antiferromagnetism state, where the magnitude of the moment is the same on every $Eu^{2+}$ ion.[28,31]

The red line in Figure 3A shows the zero-field-cooled magnetization of $CsEuCl_3$ at a field of 100 Oe with a fit to the Curie-Weiss law in white. $\theta = -3.06(1)\ K$, indicating antiferromagnetic coupling dominates like in $CsEuBr_3$. $C = 7.123(9)$ emu·K/mol, which corresponds to a moment per $Eu^{2+}$ of 7.55 $\mu_B$, deviating by 5% from the ideal spin-only moment of $Eu^{2+}$ (7.94 $\mu_B$). $\chi_0 = -1.61(2) \times 10^{-3}$ emu/mol. No magnetic phase transition is apparent between 1.8 and 300 K.

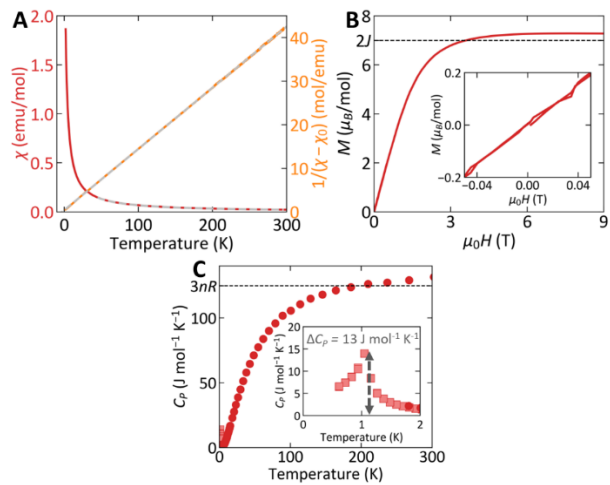

**Figure 3: $CsEuCl_3$ magnetism.** A) Temperature-dependent susceptibility at 100 Oe (red) with Curie-Weiss fit (grey) and inverse susceptibility (orange) with linear fit (grey). B) Field-dependent magnetization at 1.9 K. C) Constant-pressure heat capacity with conventional measurements shown as circles and $^3$He measurements as squares.

To further investigate the magnetic properties of $CsEuCl_3$, we measured its heat capacity from 0.6 to 300 K (Figure 3C). We find a discontinuity at 1.0 K where the heat capacity increases by is 13 J/K·mol. This change is consistent with a phase transition to an amplitude-modulated antiferromagnetic state, where the theoretical jump in heat capacity is 13.4 J/K·mol, 2/3 of what is found in the transition to an equal-moment antiferromagnetic order like in $CsEuBr_3$.[28,31] In an amplitude-modulated antiferromagnet, there is a periodic variation of the magnetic moment on each magnetic ion which is incommensurate with the lattice, and the smaller jump in specific heat is caused by the periodicity because not all spins are ordered on every ion.

The Goodenough-Kanamori rules can be used to predict and rationalize the magnetism of extended solids through superexchange interactions, which occur when a closed-shell ion (here, Cl$^-$ or Br$^-$) bridges two open-shell ions (here, Eu$^{2+}$).[29] When the two open-shell ions have the same number of unpaired electrons and the angle formed by the three atoms is close to 180°, antiferromagnetism is expected because the exchange interaction occurs through the same orbital of the closed-shell ion, and the Pauli exclusion principle requires the spins on the ions to be antiparallel to one another. If the angle is instead 90°, the exchange interaction can occur through two orthogonal *p* orbitals where the Pauli exclusion principle does not apply, so the spins will likely interact ferromagnetically.

The magnetism of both CsEuBr$_3$ and CsEuCl$_3$ can be rationalized through the Goodenough-Kanamori rules. They are both perovskites, so the Eu-X-Eu angle is close to 180° (Figure 1B-C). Both CsEuBr$_3$ and CsEuCl$_3$ should thus be antiferromagnetic, matching our experimental observations. Antiferromagnetism is consistent with what is observed in other magnetic halide perovskites such as KMF$_3$ (M = Mn, Fe, Co, Ni, Cu) and KMnCl$_3$, which all order antiferromagnetically with Néel temperatures ranging from 88-275 K.[16,18]

The Neél temperatures of CsEuBr$_3$ and CsEuCl$_3$ are similar to the temperature at which other Eu$^{2+}$-containing halides magnetically order. EuF$_2$ orders antiferromagnetically at 1.0 K,[32] EuCl$_2$ and EuI$_2$ ferromagnetically at 1.64 and 1.75 K, and EuBr$_2$ is paramagnetic with no magnetic ordering above 1.1 K.[33] EuF$_2$ is cubic and adopts the CaF$_2$ structure type, with Eu-F-Eu bond angles of 109.5°.[34] EuCl$_2$, EuBr$_2$, and EuI$_2$ are more complicated structures with multiple Eu-X-Eu bond angles in each structure that range from 82-118°;[33] their ferromagnetism is expected from Goodenough-Kanamori rules because of the near 90° bond angles.[29] The layered organic-inorganic hybrid perovskite $(C_4H_9NH_3)_2EuI_4$ does not order above 1.8 K and has a Weiss temperature of -2 K, consistent with its Eu-I-Eu bond angles near 180°.[35]

A previous report determined that nanocrystals and thermally evaporated thin films of $CsEuCl_3$ are ferromagnetic with Curie temperatures of 2.5-3.0 K using magnetic circular dichroism measurements.[25] Nanocrystalline and thin film samples may behave differently than bulk materials because of differences in preparation, interactions with the substrate, or the formation of $Eu^{3+}$ in addition to $Eu^{2+}$ to compensate for the presence of excess $Cl^-$ or deficiency of $Cs^+$. The oxide perovskite $LaMnO_3$, for example, is antiferromagnetic when prepared as a bulk material, but thin films of $LaMnO_3$ can be ferromagnetic.[26,27] The onset of ferromagnetism has been rationalized as either the result of strain in the thin film[27] or the formation of $Mn^{4+}$ to compensate for excess oxygen content.[26] We believe that ferromagnetism is unlikely in *bulk* $CsEuCl_3$ because of its negative Θ, the presence of the heat capacity anomaly at 1.0 K with a magnitude consistent with the transition to an amplitude-modulated antiferromagnetic state, and the prediction of antiferromagnetism from the Goodenough-Kanamori rules. The transition from antiferromagnetism in bulk $CsEuCl_3$ to ferromagnetism in thin film and nanocrystalline $CsEuCl_3$ demonstrates that it may be possible to tune the magnetic interactions in magnetic halide perovskites through substrate or surface effects, allowing for the creation of spintronic circuits with both ferromagnetic and antiferromagnetic components using $CsEuCl_3$ as the only magnetic material.

We next use density-functional theory calculations[36,37] to better understand the magnetic interactions in $CsEuBr_3$ and $CsEuCl_3$. Computed band structures are shown in Figures S3-S6. For both materials, the antiferromagnetic configuration is found to be slightly more stable than the ferromagnetic configuration. For CsEuBr$_3$, the antiferromagnetic configuration is 0.37 meV/f.u. more stable than the ferromagnetic configuration, and for CsEuCl$_3$, the antiferromagnetic configuration is 0.24 meV/f.u. more stable. We further analyze the results of our calculations using a simple Hamiltonian to model nearest neighbor and next nearest neighbor interactions:

$$\widehat{H} = J_1 \sum_{\langle i,j \rangle} \widehat{S}_i \cdot \widehat{S}_j + J_2 \sum_{\langle\langle i,j \rangle\rangle} \widehat{S}_i \cdot \widehat{S}_j + E_0 \qquad 2$$

where $J_1$ and $J_2$ represent the nearest and next nearest neighbor couplings, respectively, $\langle i,j \rangle$ and nearest neighbors and $\langle\langle i,j \rangle\rangle$ next nearest neighbors, and $E_0$ all nonmagnetic components of the total energy.[38,39] The results are given in Table 2; positive *J* values represent antiferromagnetic coupling and negative *J* values represent ferromagnetic coupling. We also compute the stabilizing energy of a given magnetic configuration by comparing the computed DFT energies to *E$_0$*, and these values are also presented in Table 2; negative energies indicate a given configuration is more stable than a hypothetical nonmagnetic configuration. Antiferromagnetic configurations of both materials are more stable than nonmagnetic analogs, as expected from our experimental results. A ferromagnetic configuration of CsEuBr$_3$ is 0.14 meV/f.u. less stable than hypothetical nonmagnetic CsEuBr$_3$. In contrast, a ferromagnetic configuration of CsEuCl$_3$ is 0.027 meV/f.u. *more* stable than hypothetical nonmagnetic CsEuCl$_3$. Our calculations therefore support the possibility of harnessing CsEuCl$_3$ as a switchable magnetic material because both the antiferromagnetic and ferromagnetic configurations reduce the total energy of the system compared to nonmagnetic CsEuCl$_3$. We hypothesize that very small changes in the structure of CsEuCl$_3$, such as those induced by strain in thin films or nanocrystals, cause the ferromagnetic configuration to become more stable than the antiferromagnetic configuration.

**Table 2:** Summary of computed magnetic properties. All energies are in units of meV/f.u.

|  | $J_1$ | $J_2$ | $E_{AFM}-E_0$ | $E_{FM}-E_0$ |
|---|---|---|---|---|
| $CsEuBr_3$ | 0.030 | -0.0023 | -0.22 | 0.14 |
| $CsEuCl_3$ | 0.0198 | -0.0081 | -0.26 | -0.027 |

**Conclusion**

We discover that the semiconducting halide perovskites $CsEuBr_3$ and $CsEuCl_3$ order antiferromagnetically at 2.0 K and 1.0 K. The magnitudes of the heat capacity anomalies suggest that the moments on each $Eu^{2+}$ are equal in $CsEuBr_3$ but vary periodically and incommensurately in $CsEuCl_3$; neutron diffraction measurements will be needed to confirm their magnetic structure. $CsEuBr_3$ and $CsEuCl_3$ are the first examples of bulk rare-earth halide perovskites that exhibit magnetic order. Furthermore, density-functional theory calculations show that the antiferromagnetic and ferromagnetic configurations of $CsEuBr_3$ and $CsEuCl_3$ are nearly equally stable, and both the ferromagnetic and antiferromagnetic configurations of $CsEuCl_3$ provide stabilizing interactions over a nonmagnetic analog. Our calculations combined with the previous observation of ferromagnetism in nanocrystals and thin films of $CsEuCl_3$ demonstrate that the type and degree of magnetic coupling can be tuned from antiferromagnetic to ferromagnetic through quantum confinement or substrate-film interactions.[25] The observation of magnetic ordering in bulk $CsEuBr_3$ and $CsEuCl_3$ expands the possible applications of chloride, bromide, and iodide halide perovskites beyond optical and electronic devices to now include spintronic devices where both antiferromagnetic and ferromagnetic devices can be made out of the same material.

**Methods**

*Synthesis*

CsEuBr$_3$ was synthesized by placing a stoichiometric amount of CsBr (Sigma-Aldrich, 99.999%, anhydrous) and EuBr$_2$ (Alfa Aesar, 99.99%) in a quartz tube inside an argon-filled glove box. The tube was then evacuated and flushed with argon three times before being ampouled under vacuum (10$^{-3}$ torr). The ampoule was heated in a furnace to 900 °C for 12 hours before being cooled to room temperature at a rate of 6 °C/hour and was subsequently opened in a nitrogen-filled glove box. CsEuCl$_3$ was synthesized similarly to CsEuBr$_3$ using CsCl (Alfa Aesar, 99.9%, ultra-dry) and EuCl$_2$ (Sigma-Aldrich, 99.99%).

*Characterization*

Powder X-ray diffraction measurements were performed on a Rigaku Miniflex II diffractometer using Cu kα radiation located inside a nitrogen-filled glove box. GSAS-II was used for Rietveld refinements.[38] Single crystal X-ray diffraction measurements were conducted using graphite-monochromated Mo kα radiation on a Bruker D8 Advance Eco single crystal diffractometer equipped with a Photon II detector. An Oxford Cryostream 800 flowed nitrogen gas over the sample throughout data collection. Frames were integrated using Bruker SAINT v8.40B, and a multi-scan absorption correction was applied using Bruker SADABS 2016/2. The initial solution was found using the direct method in the XS program,[40] and the structure was refined using SHELXL 2018/3[41] in the OLEX2 GUI.[42] VESTA was used to render structures.[43]

Diffuse reflectance spectra were collected using an Agilent Cary 5000 equipped with a DRA-2500 internal integrating sphere accessory on material diluted to approximately 15% w/w with dry MgO powder (Alfa Aesar, 99.95%) using Agilent powder sample holders sealed with an O-ring. Magnetic susceptibility and heat capacity data were collected with a Quantum Design EverCool II or a Dynacool PPMS. Magnetization was measured with a vibrating sample magnetometer (VSM) on powders or polycrystalline pieces of material in plastic holders. No diamagnetic corrections were applied for the material or sample holder. Heat capacity measurements used the two-tau relaxation method on polycrystalline pieces of material. Heat capacity measurements below 1.8 K on CsEuCl$_3$ used a $^3$He refrigerator.

*Computation*

Density-functional theory calculations were performed using Quantum Espresso version 7.1[36,37] using the spin-polarized generalized gradient approximation and PBE exchange-correlation functional.[44,45] Scalar-relativistic pseudopotentials from the Standard Solid State Pseudopotentials efficiency collection v1.12[46] were used, so spin-orbit coupling is not fully modeled. Specifically, the Cs pseudopotential (with 5s, 5p, 5d, 6s, and 6p as valence orbitals) is from version 1.2, and the Cl (3s and 3p as valence orbitals) and Br (4s and 4p as valence orbitals) pseudopotentials are from version 1.4 of the GBRV high-throughput ultrasoft pseudopotentials.[47] The Eu pseudopotential (with 4f, 5s, 5p, 5d, and 6s as valence orbitals) is from Topsakal and Wentzcovitch.[48] A Hubbard $U$ correction of 4.6 eV is used for the Eu 4f orbitals and 9.5 eV for the unoccupied Eu 5d orbitals as recommended by the authors of the pseudopotential. A 40 Ry kinetic energy cutoff for wavefunctions, a 320 Ry kinetic energy cutoff for charge density and potential, and a 10$^{-8}$ Ry self-consistency convergence threshold were used in all calculations. The structure of CsEuCl$_3$ used in the computations is the structure reported herein and provided in Supporting Information. The structure of CsEuBr$_3$ was downloaded from the Inorganic Crystal Structure Database[49] and was originally reported in reference [23]. The $c$ axis of CsEuCl$_3$ was doubled in all calculations except for the computation of the ferromagnetic band structure to allow antiferromagnetism to be modeled. To solve for $J_1$, $J_2$, and $E_0$ in Equation 2, we compute the total energy of spin configurations where one Eu site is spin-up and the others are spin down, in addition to the antiferromagnetic and ferromagnetic configurations. Spin polarization is restricted to the *z* direction in all calculations (*nspin = 2* command).

**Supporting Information**

The Supporting Information is available free of charge at acs.org: Band gap determination, computed band structures. (PDF)

**Accession Codes**

CCDC 2209902 contains the supplementary crystallographic data for this paper. These data can be obtained free of charge via www.ccdc.cam.ac.uk/data_request/cif, by emailing data_request@ccdc.cam.ac.uk, or by contacting The Cambridge Crystallographic Data Centre, 12 Union Road, Cambridge CB2 1EZ, UK; fax: +44 1223 336033.

**Acknowledgments**

This work is supported by the Gordon and Betty Moore Foundation as part of the EPiQS initiative under grant GBMF9066. The work at Gdańsk Tech. was supported by the National Science Centre (Poland; Grant UMO-2018/30/M/ST5/00773). Density-functional theory calculations used computational resources managed and supported by Princeton Research Computing, a consortium of groups including the Princeton Institute for Computational Science and Engineering (PICSciE) and the Office of Information Technology's High Performance Computing Center and Visualization Laboratory at Princeton University. We thank Weiwei Xie for use of the single crystal diffractometer and Stephen Zhang for helpful discussions on calculations.

**Additional Figures**

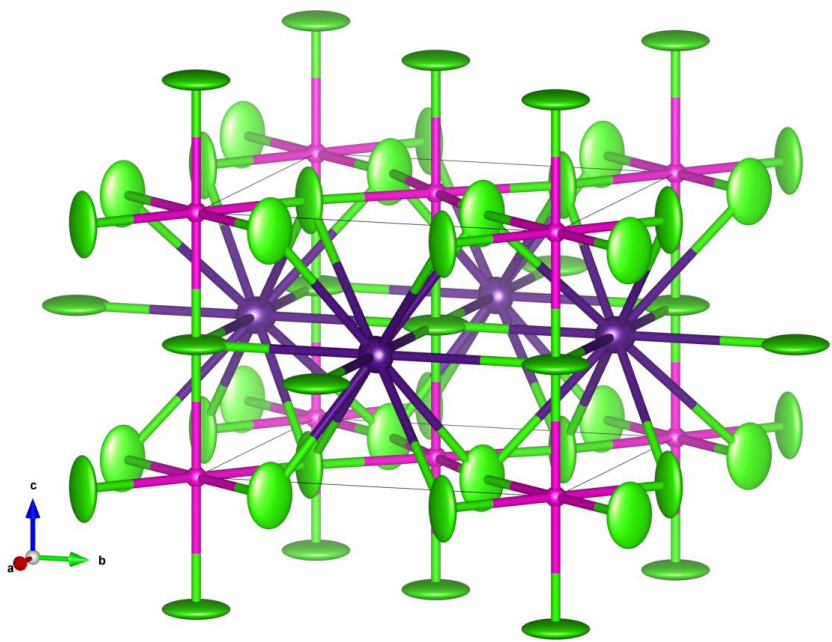

**Figure S1:** Depiction of structure of CsEuCl$_3$ with represented as 50% occupancy thermal ellipsoids. Cs is shown in purple, Eu in pink, and Cl in green.

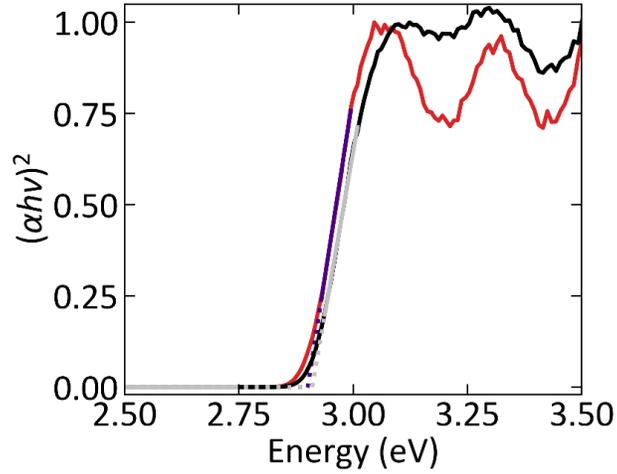

**Figure S2:** Direct bandgap Tauc plots for CsEuBr$_3$ (black, with bandgap fit in grey) and CsEuCl$_3$ (red, with bandgap fit in purple).

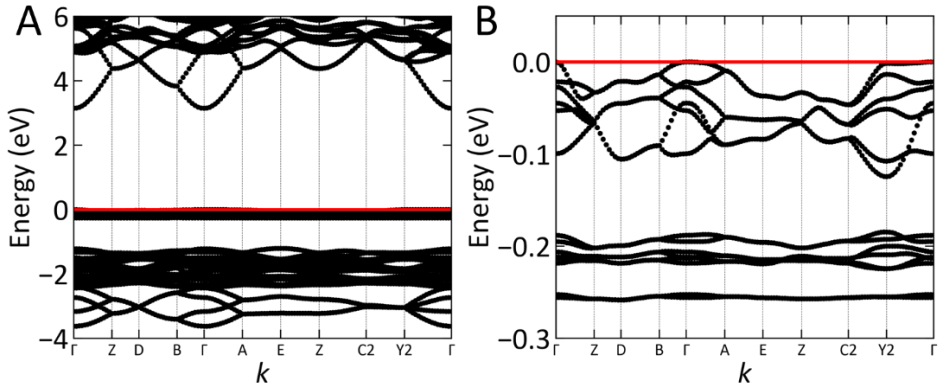

**Figure S3:** (A) Computed band structure for the spin-up electrons in the antiferromagnetic configuration of CsEuBr$_3$, space group P2$_1$/c (#14), with (B) detailed view of region near the top of the valence band. The spin-down band structure is identical.

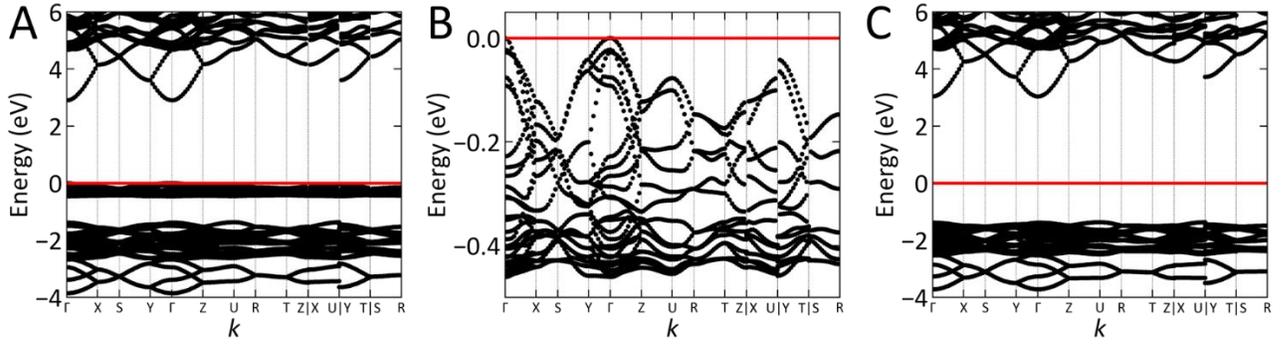

**Figure S4:** (A) Computed band structure for the spin-up electrons in the ferromagnetic configuration of CsEuBr$_3$, space group Pnma (#62), with (B) detailed view of region near the top of the valence band, and (C) for the spin-down electrons.

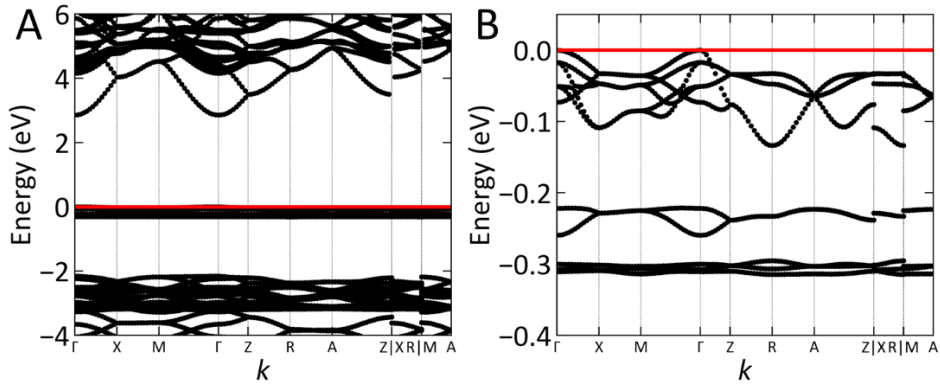

**Figure S5:** (A) Computed band structure for the spin-up electrons in the antiferromagnetic configuration of CsEuCl$_3$, space group P$_4$/mbm (#127), with (B) detailed view of region near the top of the valence band. The spin-down band structure is identical.

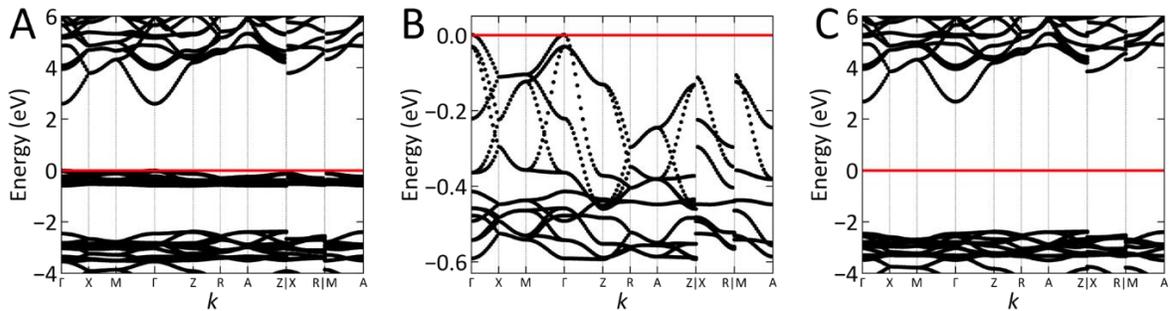

**Figure S6:** (A) Computed band structure for the spin-up electrons in the ferromagnetic configuration of CsEuCl$_3$, space group P$_4$/nmc (#128), with (B) detailed view of region near the top of the valence band, and (C) for the spin-down electrons.